\documentclass[runningheads]{llncs}
\usepackage[T1]{fontenc}
\usepackage{graphicx}
\usepackage{amsmath}
\usepackage{hyperref}
\usepackage{type1cm}
\usepackage{makeidx}
\usepackage{multicol}
\usepackage[bottom]{footmisc}
\usepackage[table,xcdraw]{xcolor}

\usepackage{newtxtext}       
\usepackage[varvw]{newtxmath}   
\usepackage{framed}
\usepackage{xcolor}
\usepackage{quoting,xparse}

\definecolor{formalshade}{rgb}{0.95,0.95,1}
\definecolor{darkblue}{rgb}{0.145, 0.118, 0.580}

\newenvironment{formal}{%
  \MakeFramed{\advance\hsize-\width\FrameRestore}%
  \noindent\hspace{-4.55pt}

  \vspace{2pt}\vspace{2pt}%
}
{%
  \vspace{2pt}\endMakeFramed%
}

\begin{document}

\title{PRISM: A Personalized, Rapid, and Immersive Skill Mastery framework for personalizing experiential learning through Generative AI}
\titlerunning{PRISM: A Personalized, Rapid, and Immersive Skill Mastery framework} 

\author{
Yu-Zheng Lin\inst{1} \and
Karan Patel\inst{2} \and
Ahmed Hussain J Alhamadah\inst{2} \and
Bono Po-Jen Shih\inst{4} \and
Matthew William Redondo\inst{2} \and
David Rafael Vidal Corona\inst{3} \and
Banafsheh Saber Latibari\inst{1} \and
Jesus Pacheco\inst{3} \and
Soheil Salehi\inst{1} \and
Pratik Satam\inst{2}\thanks{Corresponding Author.}
}

\authorrunning{Lin et al.}

\institute{
Department of Electrical and Computer Engineering, University of Arizona, Tucson, USA\\
\email{\{yuzhenglin, banafsheh, ssalehi\}@arizona.edu}
\and
Department of Systems and Industrial Engineering, University of Arizona, Tucson, USA\\
\email{\{karanpatel, alhamadah, mredondo245, pratiksatam\}@arizona.edu}
\and
Department of Industrial Engineering, University of Sonora, Hermosillo, Mexico\\
\email{\{david.vidal, jesus.pacheco\}@unison.mx}
\and
Leonhard Center for Enhancement of Engineering Education, The Pennsylvania State University, University Park, USA\\
\email{bps5848@psu.edu}
}

\maketitle

\begin{abstract}
Artificial intelligence, especially with the emergence of generative AI (gen-AI), a subset of AI that focuses on creating new content, is driving a transformative revolution across all industries. This evolution opens new avenues for training and education, creating opportunities to upskill, adapt, and modify curriculum, coursework, and training to emerging demands, keeping pace with rapid technological changes while ensuring learning efficiency, scalability, and suitability for distance learning. This book chapter proposes \textbf{\textit{PRISM:}} \textbf{P}ersonalized, \textbf{R}apid, and \textbf{I}mmersive \textbf{S}kill \textbf{M}astery, a scalable framework that explores the use of gen-AI, to personalize experiential learning for each trainee's learning needs through Digital Twins and trainee sentiment analysis. This book chapter also introduces a Multi-Fidelity Digital Twin for Education (MFDT-E) framework, a scalable framework that maps Digital Twin design requirements to Bloom's Taxonomy, and the Kirkpatrick model to access learning outcomes at different levels. The MFDT-E framework quantifies digital twins' (DT) design requirements through their use of fidelity, classifying them broadly into three separate levels-  low, medium, and high — corresponding to undergraduate, master's, and doctoral education stages. The MFDT-E, then integrated into the PRISM framework, allows personalized experiential learning, where the trainee performs experiential learning exercises through interaction with the DT, while the training is personalized through the use of Gen-AI-based sentiment analysis to measure student knowledge comprehension instep with Bloom's taxonomy, and usage of Retrieval-Augmented Generation (RAG) to adapt and generate new content when student comprehension drops below the targets defined in the courses learning outcomes. This study demonstrates the effectiveness of the PRISM framework by combining generative AI and multi-fidelity digital twins. GPT-4 achieved a 91\% F1 score in zero-shot sentiment analysis of teacher-student dialogues, while GPT-3.5 reached 79.5\% accuracy in internet slang scenarios, outperforming traditional models. A qualitative-to-quantitative sentiment approach revealed emotional dynamics during learning, supporting personalized feedback. Four VR modules—Home Scene, Factory Floor Tour, Capping Station Digital Twin, and PPE Inspection Training—were developed to deliver immersive, low-fidelity Industry 4.0 training, highlighting the system's scalability and educational value.
\end{abstract}

\section{Introduction} \label{}

The growth of Artificial Intelligence (AI) has revolutionized the world, transforming all aspects of modern life, including all topics in electrical engineering, computer engineering, industrial engineering, systems engineering, and mechanical engineering. This transformation has not only changed the topics that need to be taught to the students but also changed the scale of teaching, with an ever-present need to modernize the curriculum to incorporate the rapidly evolving learning needs of the industry while ensuring compliance with traditional curriculum certification frameworks like ABET. 
One example of AI transforming learning and training needs is the rapid adoption of automated systems, like smart manufacturing plants and autonomous vehicles, which are changing today's engineering workforce needs, increasing skillset gaps, especially for the older workforce. With an increasing emphasis on STEM skillsets like robotics, automation, artificial intelligence (AI), and security, the workforce must be reskilled and upskilled to meet future industry needs while at the same time training the new workforce through realistic and hands-on experiential learning. However, such hands-on experiential learning requires access to specialized hardware that is not easily accessible, is expensive, has a huge learning curve, and has training limitations due to the risk of personnel or equipment damage. In addition to this, such workforce development programs have to train large groups of people globally from vastly different backgrounds. Trainees/Students from such different backgrounds have different learning styles depending on their diverse upbringings, educational backgrounds, and motivations, all contributing to their successful program completion. Successful program completion is especially challenging for students from marginalized, underrepresented communities, as a historical lack of access to high-quality education during their formative years (pre/middle/high schools) impedes their success.

To address these challenges, there is a pressing need for innovative solutions that provide scalable, safe, and personalized hands-on training environments. One promising approach is using virtual and augmented learning tools that simulate real-world engineering systems. Among these tools, Digital Twin (DT) technology is a transformative solution bridging the gap between physical hardware and digital learning environments. DTs were initially conceptualized in aerospace as a digital counterpart to physical systems, facilitates real-time interaction between virtual models and their physical counterparts through continuous status updates. Driven by the increasing demand for advanced monitoring, diagnostics, and predictive analytics in industrial systems and the advancements brought by Industry 4.0, DTs have emerged as a key enabler of complex real-time behavior modeling. By integrating data from diverse sensors and IoT devices, these DTs accurately reflect physical systems while enhancing predictive capabilities, enabling industrial system optimization, fault detection, and lifecycle management. Furthermore, integrating machine learning and artificial intelligence has significantly enhanced the adaptability of DTs, improving the accuracy of behavior models under varying conditions. As a result, their applications across manufacturing, aerospace, and energy sectors have improved efficiency, reliability, and cost-effectiveness \cite{soori2023digital}.

Given these capabilities, DTs are increasingly being used in educational settings as immersive and interactive training platforms. When integrated into online courses, they allow learners to engage in realistic simulations of engineering systems—without needing physical access to expensive or hazardous equipment. This makes DTs particularly valuable for distance learning and remote workforce development, offering a safe, cost-effective, and scalable alternative to traditional lab-based instruction. Furthermore, by combining the flexibility of online education with the experiential nature of hands-on training, DT–based learning environments can improve engagement, motivation, and learning outcomes. This is especially beneficial for working professionals juggling jobs and family commitments, as well as students from underrepresented or under-resourced backgrounds who may lack access to physical labs or training facilities. In this way, DTs not only addresses the limitations of traditional online education but also promotes broader accessibility and equity in technical training programs.

Although DTs provides an effective and engaging simulation environment, the presence of a realistic training platform alone is not sufficient to ensure successful learning outcomes. Learners, especially those from non-traditional backgrounds, still require guidance, contextual understanding, and timely feedback to make the most of these experiences \cite{banks2019mitigating}. Traditionally, this role is filled by skilled instructors who can adapt their teaching to the individual needs of the learners. However, with the scale and diversity of online learners, providing such personalized instruction becomes a significant challenge \cite{jaggars2011online}. This is where Large Language Models (LLMs) offer a transformative opportunity. By leveraging recent advancements in natural language processing and AI, LLMs can serve as intelligent, always-available teaching assistants that help bridge the instructional gap in online and remote learning. They can provide context-sensitive explanations, adaptive tutoring, just-in-time hints, and even simulate expert feedback, all tailored to the learner's background, pace, and current level of understanding. To this end, we propose the PRISM framework (Personalized Rapid and Immersive Skill Mastery). PRISM integrates DT technology with LLM-powered instructional agents to provide scalable, personalized, and inclusive technical training. Within PRISM, learners interact not only with a simulated engineering system via a DT, but also receive intelligent support through an embedded AI tutor, creating a comprehensive and self-guided learning ecosystem that aligns with the principles of being as helpful, honest, and harmless (HHH) as possible \cite{askell2021hhh}.

The main contributions of this chapter are as follows:
\begin{itemize}
   \item \textbf{PRISM Framework:}
Proposed the PRISM (Personalized, Rapid, and Immersive Skill Mastery) framework, which leverages Generative AI and DT technology to enable personalized learning. This framework integrates the analysis of the sentiment of learners with the generation of personalized content, closely aligning the learning experience with individual needs.
    \item \textbf{Multi-Fidelity DT Educational (MFDT-E) Framework:}
Developed the Multi-Fidelity DT Educational (MFDT-E) framework, combining DT design requirements with Bloom's Taxonomy and the Kirkpatrick evaluation model. By leveraging a more accurate and honest representation of learning stages and progression, this systematic approach effectively assesses and enhances learning outcomes at various educational levels (undergraduate, master's, doctoral).
    \item \textbf{LLM-based Zero-Shot Sentiment Analysis Approach:}
Developed a zero-shot sentiment analysis method based on large language models (e.g., GPT-4, GPT-3.5), successfully applied to teacher-student dialogues and internet slang scenarios, demonstrating high accuracy on classification tasks. Moreover, this technique effectively captures real-time learner emotional dynamics and converts qualitative sentiments into quantitative indicators, enabling a helpful and precise feedback mechanisms.
    \item \textbf{Development and Implementation of Virtual Reality (VR) Modules:}
Created and validated four educational modules based on VR and low-fidelity DTs (Home Scene, Factory Floor Tour, Capping Station DT, and Personal Protective Equipment Inspection Training). These modules showcase the scalability of the PRISM framework, providing a safe, immersive, and efficient Industry 4.0 educational platform. They effectively overcome traditional limitations such as high equipment costs and harmful risks, significantly enhancing practicality and accessibility in distance learning.
\end{itemize}

The rest of the chapter is organized as follows: In Section \ref{sec:related_work}, we discuss related works, include DT, DT in eductation application, and LLMs in education application; in Section \ref{sec:main_framework}, we present the PRISM framework with multi-fidelity DT for education framework; in Section \ref{sec:experiments}, we present the experimental evaluation of the PRISM framework along with the zero-shot sentiment analysis and multi-fidelity DT labs, and in Section \ref{sec:conclusion}, we conclude the chapter.

\begin{figure*}[!t]
  \centering
    \includegraphics[width=1\linewidth]{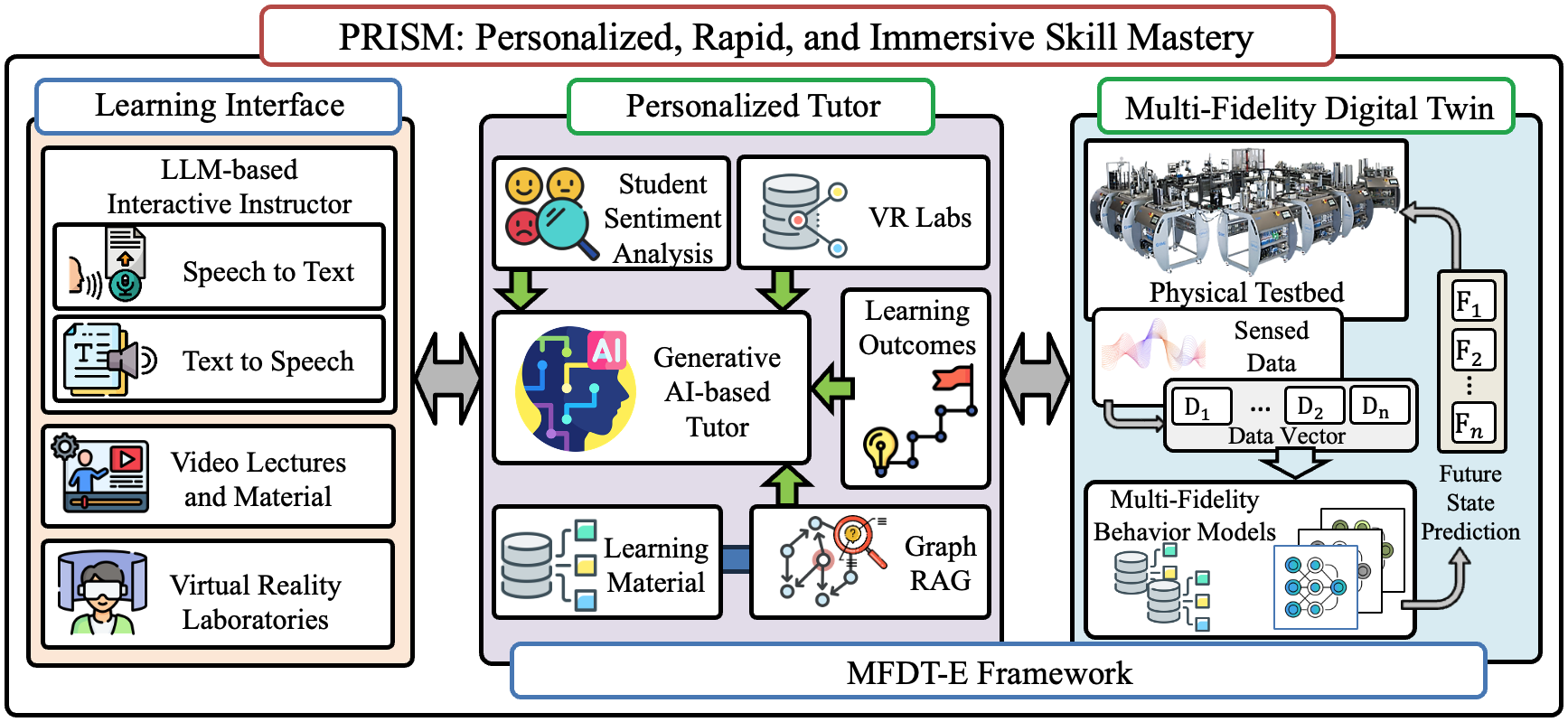}
    \caption{PRISM: A Personalized, Rapid, and Immersive
Skill Mastery framework for personalizing experiential learning through Generative AI}
    \label{fig:framework}
\end{figure*}

\section{Related Work} \label{sec:related_work}
In this section, we review three key areas relevant to our study: DTs, their applications in industrial education, and the role of large language models (LLMs) in educational settings. DT technology has evolved from its initial aerospace applications to become a critical tool in various industries, enabling real-time system monitoring, predictive maintenance, and operational optimization. Integrating DTs into industrial education has been explored to improve workforce training, offering immersive and interactive learning experiences while addressing the challenges of cost and complexity in industrial environments \cite{eriksson2022applying,liljaniemi2020using,longo2023prepare}. Several studies have demonstrated the potential of DTs in education, from web-based training platforms to virtual reality-enhanced learning experiences \cite{lei2023web,bucchiarone2022gamification}. However, However, systematically evaluating their impact on educational outcomes remains an open challenge. Meanwhile, recent advancements in LLMs have introduced new opportunities for educational applications, including intelligent tutoring systems, automated assessment tools, and personalized learning experiences. By reviewing these three interconnected areas, we establish a foundation for understanding how DTs and LLMs can be leveraged to enhance education and improve learning outcomes.

\subsection{Digital Twin}
Digital twin (DT) technology originated in aerospace engineering and was formally proposed by NASA in 2010 to improve the design, monitoring, and maintenance of complex space vehicle systems \cite{glaessgen2012digital}. Rooted in systems engineering and simulation, the concept of DTs differs from traditional offline simulations in that it incorporates real-time sensor updates and historical data alongside physical models. This integration enables the DT to reflect highly realistic behaviors of its corresponding physical system \cite{shafto2010draft}. To enhance accessibility and usability, modern DTs not only simulate physical behavior but also emphasize user-friendly visualization. Constructing a visual representation of the physical system and translating it into an intuitive, interactive interface has become a key development challenge in DT technology \cite{tao2022digital}.

The development of accurate behavioral models and the design of intuitive user interfaces have become central challenges in DT (DT) research. Traditional numerical modeling techniques often struggle to capture the intricacies of complex systems. However, recent advances in machine learning and computational power have positioned data-driven modeling as a practical and increasingly favored alternative \cite{rathore2021role,min2019machine,lin2023dt4i4}. By reducing the complexity associated with DT development, machine learning has significantly broadened the scope of DT applications across various domains \cite{tao2018digital}. The versatility of DTs has led to its integration into a broad spectrum of disciplines. In climate science, DTs contribute to more accurate weather forecasting and enhance disaster response strategies by simulating intricate atmospheric dynamics \cite{rao2023developing}. In the medical domain, DTs facilitate the advancement of precision medicine by enabling individualized diagnostics and treatment planning through patient-specific computational models \cite{coorey2021health}. In the manufacturing sector, DTs function as dynamic virtual counterparts of physical production systems, supporting real-time monitoring, predictive maintenance, process optimization, and safety assurance \cite{cimino2019review}. Collectively, these implementations highlight the transformative capabilities of DTs and underscore their growing relevance across diverse application areas.

\subsection{Digital Twin in Education Application}
In education, the complexity and cost of engineering environment limitations cause the challenge of workforce training. Using DTs for industrial education to respond to workforce development needs has become a potential option and has attracted attention \cite{hazrat2023developing}. DTs can be used as a training tool, allowing learners to conduct personalized training through the virtual and real combination of DTs and factory environments, and can greatly improve resource optimization and operational efficiency \cite{lu2020digital}. Lei et al. proposed a web-based DT training environment for communication maintenance and power systems training\cite{lei2023web}. They used optical fiber fault repair as an example to teach students how to understand the actual before and after repair situations through the DT. Moreover, Martínez-Gutiérrez et al. proposed in 2023 the use of virtual reality combined with DTs to train operators to operate mobile robots \cite{martinez2023convergence}. The results explained that DTs can be combined with actual equipment to become a compromise between cost and teaching effectiveness. In process engineering experimental education, Boettcher et al. implemented a jet pump in virtual reality through DT technology, allowing students to solve problems independently in complex and sometimes vague scenarios \cite{boettcher2023developing}.

These examples illustrate the potential for DTs in education and workforce development across different. However, current research on DTs in education still lacks systematic evaluation methods and potential value to higher education. This study introduces Bloom's Taxonomy \cite{forehand2010bloom} into the digital learning framework to systematically evaluate learning performance. In terms of education, Awouda et al. used this method to evaluate the effectiveness of open-source IoT in education \cite{awouda2024bloom}. Additionally, Gajek et al. used this method to evaluate the learning effectiveness of process safety \cite{gajek2022process}. These examples have educational objectives related to Industry 4.0 and illustrate that Bloom's Taxonomy is a suitable option for evaluating the effectiveness of industrial education.

\subsection{Large Language Models in Education Application}
\textit{``The term \textit{generative AI} refers to computational techniques that are capable of generating seemingly new, meaningful content such as text, images, or audio from training data \cite{feuerriegel2024generative}.''} Large language models (LLMs), as a prominent subclass of generative AI, specialize in processing and producing human language with high contextual relevance. These models leverage extensive training on diverse textual corpora to support complex reasoning, dialogue, and content generation tasks \cite{fui2023generative}.

With the rapid advancement of Large Language Models (LLMs) in the field of natural language processing, their applications in education have attracted growing interest \cite{kasneci2023chatgpt,osunbunmi2024board}. Due to their capabilities in generative language processing, contextual understanding, and knowledge reasoning, LLMs have been explored for use in intelligent tutoring systems, personalized learning environments, and automated content generation \cite{moore2023empowering}. For example, Zhang et al. propose SimClass\cite{zhang2024simulating}, a virtual classroom framework that integrates LLMs with multi-agent systems to simulate interactions between teachers, students, and peers. In an experiment involving more than 400 students, SimClass was evaluated using the Flanders Interaction Analysis System and the Community of Inquiry framework. The results indicate that the system can improve student engagement and support better learning outcomes, suggesting its potential in educational applications. In addition, Chiang et al. propose LLM evaluation as a reproducible and cost-effective alternative to traditional human evaluation in natural language processing\cite{chiang2023can}. They test LLMs by feeding them exact instructions and inputs as human evaluators and extracting Likert scale ratings from model outputs. Their findings show that advanced LLMs produce highly correlated evaluations from expert English teachers, especially in relevance and fluency assessments. Furthermore, LLM evaluations exhibit low variance between instructions and sampling settings, suggesting robustness and consistency. Although not merely replacing human judgment, the study positions LLMs as a promising complementary tool for efficient and scalable evaluation of NLP systems. Tsai et al. explored the integration of large language models (LLMs), specifically Chat-GPT, into chemical engineering education to improve students' problem-solving abilities \cite{tsai2023exploring}. They designed and implemented an experimental course that involved 29 students (from the freshman to the graduate level). LLMs were used to build virtual models for core chemical engineering problems such as thermodynamics, reaction engineering, and mass transfer. The study demonstrated that LLMs could significantly improve problem solving efficiency and conceptual understanding, particularly in tasks involving code generation and scenario simulation. 

These studies collectively highlight the transformative potential of LLMs in reshaping educational practices across disciplines. By simulating realistic classroom dynamics, offering scalable and consistent evaluations, and enhancing domain-specific learning, LLMs supplement traditional instruction and open new pathways for personalized, interactive, and adaptive education. 

\section{PRISM - Personalized Rapid and Immersive Skill Mastery} \label{sec:main_framework}
In this section, we introduce the core concepts of the PRISM (Personalized Rapid and Immersive Skill Mastery) framework, which integrates multi-fidelity DTs with generative AI as a tutor to achieve rapid, personalized, and immersive skill development.

\subsection{Multi-Fidelity Digital Twin for Education Framework}
\begin{figure*}[!b]
  \centering
    \includegraphics[width=1\linewidth]{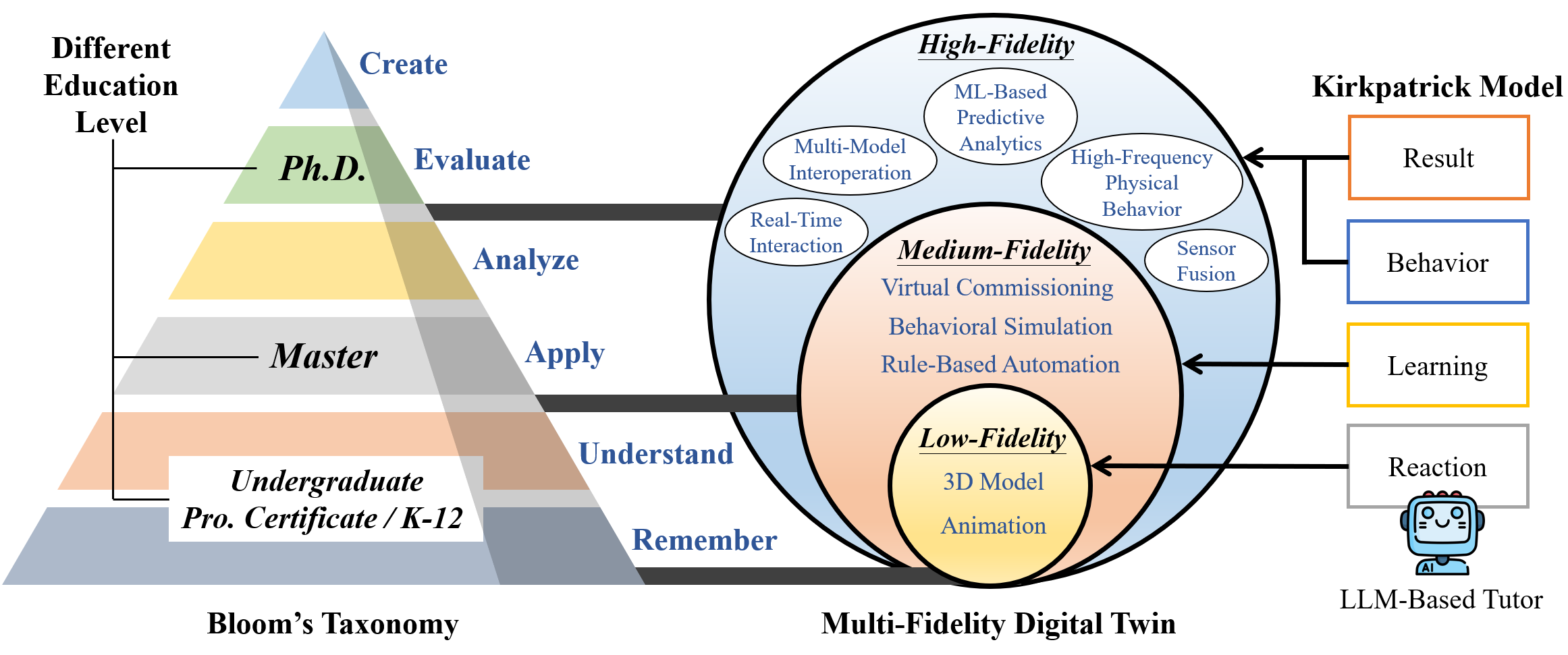}
    \caption{Educational Learning Stages and Outcome Mapping Based on a Multi-Fidelity Digital Twin Model}
    \label{fig:mfdt-e}
\end{figure*}

The Multi-Fidelity Digital Twin for Education Framework (MFDT-E) is an educational framework grounded in the concept of Multi-Fidelity Digital Twins, which integrates multiple educational levels with Bloom's Taxonomy and the Kirkpatrick evaluation model (see Figure \ref{fig:mfdt-e}) \cite{lin2025personalized,krathwohl2002revision,smidt2009kirkpatrick}. This framework interweaves these four components to form a comprehensive structure for both learning and assessment. By leveraging DT technology, it aims to fulfill industrial education objectives while guiding learners from foundational understanding to creative application, all while addressing measurable learning outcomes and behavioral change.

In this framework, the six levels of the cognitive domain of Bloom's Taxonomy are \textit{Remembe}, \textit{Understanding}, \textit{Application}, \textit{Analysis}, \textit{Evaluate}, and \textit{Create}, which correspond to different stages of education -- \textit{Undergraduate}, \textit{Master}, and \textit{Doctoral} degrees. As learners progress through the educational stage, their learning goals shift from memorizing and understanding basic knowledge to higher-level like analysis, evaluation, and creation. In the educational needs of different stages, multi-fidelity DTs can become an educational tool, providing reasonable learning assistance for different educational levels and making educational goals at different stages successful.

The integration of multi-fidelity digital twin (DT) technology into educational settings demonstrates strong alignment with established educational objectives, particularly those defined by Bloom’s taxonomy. At the foundational level, low-fidelity DTs—typically composed of simple three-dimensional models and basic animations—serve as effective tools for introductory education. These models assist learners in grasping fundamental concepts and cultivating initial interest in the subject matter. Such applications correspond to the lower tiers of Bloom’s taxonomy, namely \textit{Remember} (retrieving relevant knowledge from long-term memory) and \textit{Understand} (interpreting the meaning of instructional content, including oral, written, and graphical communication). In more advanced educational contexts, such as at the Master’s level, medium-fidelity DTs offer capabilities such as virtual commissioning, behavioral simulation, and rule-based automation. These afford learners the opportunity to engage with more complex scenarios, facilitating deeper cognitive engagement. These applications reflect the intermediate cognitive processes of Bloom’s taxonomy, particularly \textit{Apply} (executing procedures in defined contexts) and \textit{Analyze} (decomposing information into components and understanding their interrelationships within a structural framework). At the doctoral level, high-fidelity DTs incorporate advanced functionalities, including real-time system interaction, predictive analytics powered by machine learning, and multi-model interoperability. These features support higher-order cognitive activities such as creative problem-solving and critical evaluation. As such, they align with the highest levels of Bloom’s taxonomy—\textit{Evaluate} (formulating judgments based on defined criteria and standards) and \textit{Create} (assembling components to produce innovative and coherent outcomes).

In evaluating students' learning effectiveness, we introduce the Kirkpatrick model \cite{smidt2009kirkpatrick} into the framework, which can ensure a more comprehensive and and honest evaluation of students' learning effectiveness at different educational levels. At the undergraduate level, using low-fidelity DTs, the emphasis is on the first level: reaction, focusing on how students perceive the training, its relevance to their future careers, and their participation in the learning process. As students advance to the Master's degree, Level 2: Learning becomes the focus, assessing how students acquire expected knowledge, skills, and attitudes through mid-fidelity DTs, thereby facilitating deeper application and analysis. At the doctoral level, using high-fidelity DTs, the framework moves to Level 3: Behaviors and Level 4: Result. The focus here is on how students apply their knowledge and skills in real-world settings to drive organizational results and innovation. This integration can help educators gain a comprehensive view of learner progress. 

\subsubsection{Low Fidelity DT for Undergraduate, Professional Certificate and K-12}
At the undergraduate level, educational emphasis is placed on building foundational knowledge and cultivating essential technical skills, equipping students with the basic competencies required for effective problem-solving \cite{koen1998abet}. In this context, low-fidelity DTs can offer simplified representations of industrial systems that facilitate conceptual understanding. While low-fidelity DTs cannot simulate precise system behavior, these models are adequate for tasks involving the core principles of \textit{``remembering''} and \textit{``understanding.''} Low-fidelity DTs can enable students to engage in fundamental aspects of system operation without being overwhelmed by the complexity of high-fidelity simulations, causing additional learning confusion \cite{russell1984effects}. Features of these models may include basic manufacturing processes, simplified machine operations, or basic control systems that are sufficient for introductory courses in engineering and related disciplines.

\subsubsection{Medium Fidelity DT for Master Degree}
As students advance to the postgraduate level, especially master's programs, the educational focus shifts to \textit{``applied''} and \textit{``analytical''} knowledge \cite{cranch1994next,backa2014future}. At the master's level, medium-fidelity DTs become more appropriate. These models offer a higher level of detail and interactivity, enabling students to solve complex problems and analyze real-world scenarios with greater precision \cite{choe2020master}. Medium-fidelity DTs are particularly suitable for project-based learning, by modeling realistic system behavior and incorporating rule-based logic or behavioral models, these DTs support the development of applied engineering skills and promote deeper engagement with domain-specific content. As a result, students reinforce their theoretical understanding and enhance their ability to address industry-relevant problems through iterative experimentation and decision-making. The ability to manipulate variables, test hypotheses, and analyze results in a controlled digital environment is consistent with the cognitive processes of application and analysis described in Bloom's taxonomy. Medium-fidelity DTs can simulate manufacturing processes with detailed parameters that students can adjust to optimize output or resolve potential issues, promoting a deeper understanding of system dynamics and interdependencies.

\subsubsection{High Fidelity DT for Doctoral Degree}
At the doctoral level, the educational focus is on \textit{``creating''} and \textit{``evaluating''} new knowledge, requiring the highest degree of cognitive engagement. High-fidelity DTs provide the detailed, accurate, and dynamic modeling needed for advanced research and innovation at the doctoral level. These models enable doctoral students to design and test new systems, conduct complex experiments, and evaluate the feasibility of novel solutions in virtual environments that closely reflect real-world conditions \cite{buswell2021purpose}. Integrating real-time data, advanced physics-based modeling, and artificial intelligence-driven analytics in high-fidelity DT supports the rigorous requirements of doctoral research, allowing for exploring cutting-edge technologies and developing new methods.

\subsection{Leveraging Generative AI-based Technologies for DT-based Personalized Learning}
Integrating a virtual tutor based on generative AI into a DT learning environment offers new possibilities for adaptive and responsive education. Within this setting, the virtual tutor can leverage the capabilities of large language models to provide real-time, personalized guidance tailored to individual learner needs. By continuously observing learner interactions and the simulated learning context maintained by the DT, the system can assess performance, deliver targeted feedback, suggest helpful learning strategies, and address knowledge gaps. This integration supports a more data-driven and dynamic approach to instruction, aligning educational support with the learner’s progress in real-time.

\begin{figure*}[!b]
  \centering
  \includegraphics[width=\linewidth]{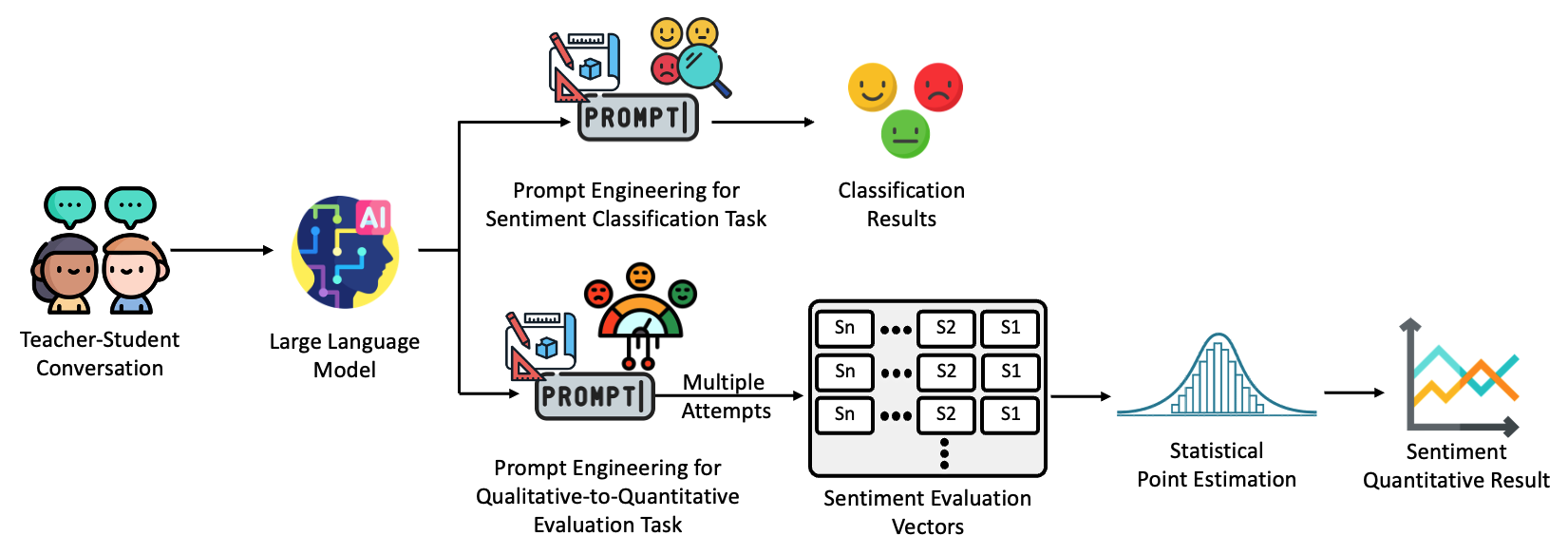}
    \caption{Zero-shot LLM-based sentiment analysis pipeline}
    \label{fig:llm_sentiment}
\end{figure*}

Integrating large language models (LLMs) as virtual tutors could revolutionize the evaluation of the Kirkpatrick model by leveraging sentiment analysis and other natural language processing (NLP) techniques to provide real-time, data-driven insights into learning effectiveness \cite{lin2025personalized}. At Level 1 (Reaction), LLMs could analyze student feedback surveys, discussion forums, and course evaluations using sentiment analysis to gauge engagement, satisfaction, and common areas of concern. By identifying emotional tone, recurring keywords, and sentiment trends, educators could refine instructional methods and address pain points dynamically. At Level 2 (Learning), LLMs could assess students’ understanding by analyzing responses in quizzes, assignments, and interactive discussions. Sentiment analysis combined with text-based cognitive evaluation could reveal confidence levels, areas of uncertainty, and conceptual gaps, enabling personalized feedback and adaptive learning pathways. At Level 3 (Behavior), LLMs could track students' ability to apply knowledge in real-world contexts by analyzing their written reflections, project reports and peer interaction \cite{chiang2024large,chan2023chateval}. Changes in sentiment over time, along with NLP-based assessments of reasoning and critical thinking, could provide insights into how well learners translate theoretical knowledge into practical actions. At Level 4 (Results), LLMs could monitor long-term impacts by evaluating professional reports and research contributions\cite{zhou2024llm}. Sentiment trends, combined with semantic analysis of applied skills, could help determine whether educational experiences lead to tangible improvements in job performance, innovation, or problem-solving. By integrating sentiment analysis across all four levels of the Kirkpatrick model, LLMs could enable a more adaptive, data-driven approach to educational assessment, helping institutions refine curriculum design, personalize learning experiences, and ensure that knowledge acquisition translates into meaningful real-world outcomes.

For sentiment analysis, this study uses LLM-based zero-shot sentiment analysis, as shown in Figure \ref{fig:llm_sentiment}, which has the advantage of not relying on a large amount of labeled data in a specific field and can be quickly applied to new tasks or different scenarios, significantly reducing development and training costs. Through well-designed prompt engineering, we enable the model to understand the task context and generate reasonable results accurately \cite{white2023prompt}. We divide the task into classification tasks and qualitative-to-quantitative evaluation tasks in sentiment analysis. The classification task mainly analyzes the polarity of sentiment, such as ``positive'' or ``negative.'' The model evaluates the learner's sentiment state by examining the learner's input text and makes a comprehensive judgment based on other parameters of the learning environment. The qualitative-to-quantitative task converts the sentiment into specific quantitative indicators, such as expressing the positive and negative of the learner's sentiment in numerical form to assess the mutual influence of teacher-student interaction during the dialogue process.

Based on this sentiment analysis framework, Retrieval-Augmented Generation (RAG) \cite{lewis2020retrieval} is incorporated to further refine the personalized responses of the virtual tutor. By dynamically retrieving relevant domain-specific knowledge based on the sentiment and contextual cues in learner interactions, RAG ensures that instructional feedback is accurate and adaptive to the student's emotional and cognitive state. This approach enables the virtual tutor to tailor explanations, troubleshooting guidance, and real-time learning support, addressing conceptual misunderstandings and engagement barriers. By integrating sentiment-aware retrieval mechanisms, LLM-based tutors can effectively bridge knowledge gaps and optimize instructional strategies, fostering a more responsive and personalized learning experience.

\subsection{LLM-based Emotion-Aware Enhanced Guidance}
To effectively support students conducting experiments in a Digital Twin (DT)-based learning environment, it is essential to deliver personalized and adaptive guidance that dynamically reflects each learner’s cognitive and emotional states. As students engage with interactive systems, their performance and progress are influenced not only by their understanding of the subject matter but also by affective factors such as motivation, frustration, or confidence \cite{kechaou2011improving,li2022key,dehbozorgi2020sentiment}. Thus, the ability to continuously monitor behavioral data from the DT environment while simultaneously inferring emotional context through natural interaction is crucial for making an interactive AI system helpful. Such a system enables emotion-aware, context-sensitive instructional support, which fosters deeper engagement and improved learning outcomes.

\begin{figure*}[b!]
  \centering
    \includegraphics[width=0.8\linewidth]{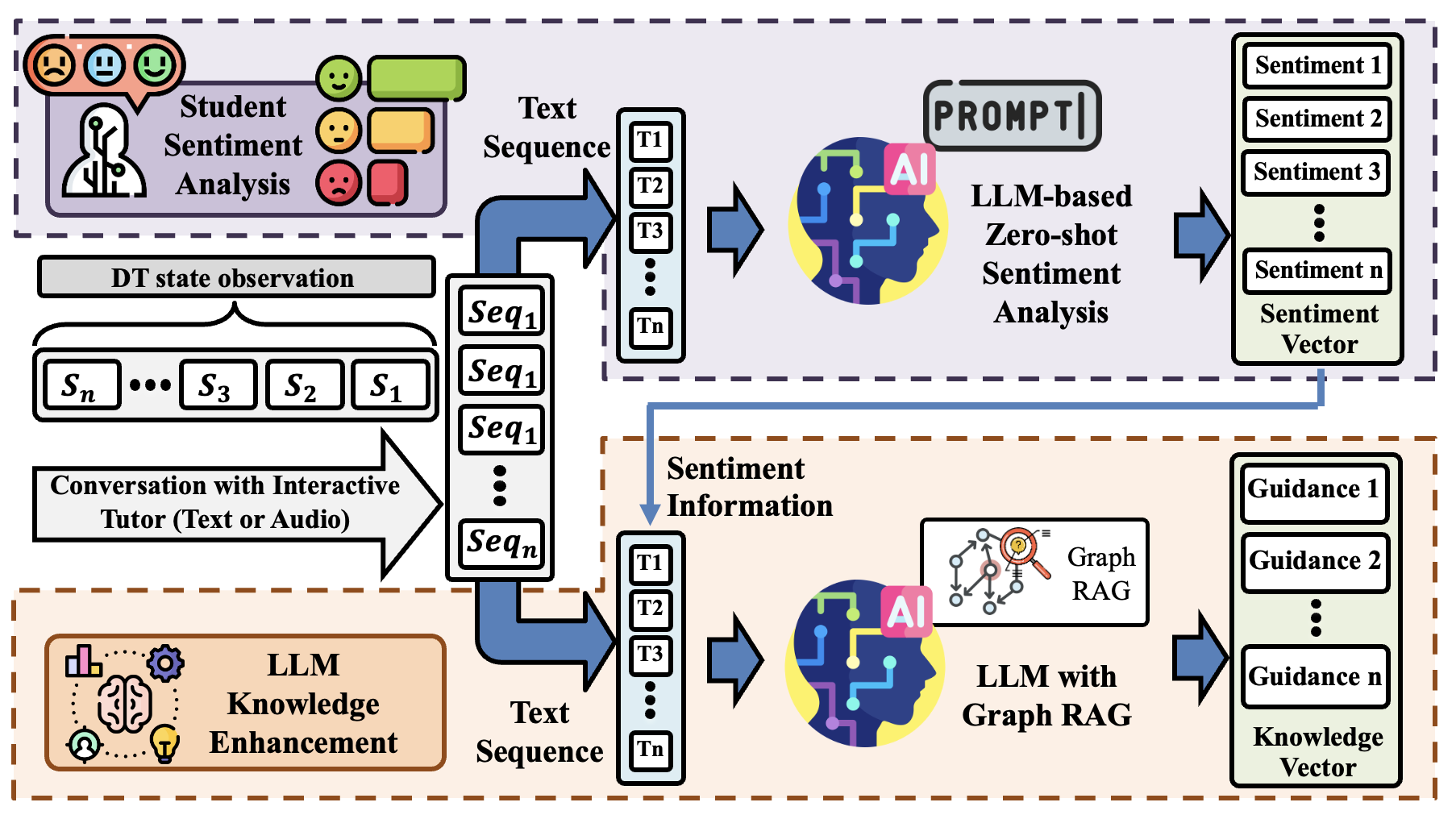}
    \caption{LLM-based Emotion-Aware Enhanced Guidance Architecture}
    \label{fig:EAEG}
\end{figure*}

To address this need, we'd like to propose the Emotion-Aware Enhanced Guidance Framework, which integrates Student Sentiment Analysis and Knowledge-Augmented Guidance Generation into a cohesive architecture. This architecture is designed to deliver intelligent tutoring by leveraging the capabilities of LLMs in tandem with sentiment-aware mechanisms. By jointly considering DT states and human behavior, the system adapts its instructional strategies to the different needs of each learner, thereby enhancing the overall interactivity and responsiveness of the educational platform.

The architecture is shown in Figure \ref{fig:EAEG}. During conversations with an Interactive Tutor, whether through text or speech, the system captures the dialogue content and state observations from the DT environment (e.g., learning activities and task completion patterns). These are processed into textual sequences that serve as input for a multistage pipeline. In the first stage, a zero-shot sentiment analysis model, powered by an LLM, interprets the emotional undertones of the student's responses and generates a sentiment vector. This vector is then passed to the second stage, where a graph-based retrieval-augmented generation (Graph RAG) module, which enhances LLM with structured domain knowledge, produces tailored instructional guidance. The result is a set of pedagogically informed recommendations encapsulated in a knowledge vector, allowing precise real-time adjustments to the student’s learning pathway based on their cognitive and emotional needs.

\section{Experiments} \label{sec:experiments}
In this chapter, we conduct a series of experiments to evaluate the potential and effectiveness of applying generative AI within the PRISM framework for educational purposes. These experiments include zero-shot sentiment analysis, enhancing LLM expertise with GraphRAG, and leveraging DT technology to develop lab modules that strengthen overall instructional applications.
\subsection{LLM-based Zero-Shot Sentiment Analysis} 
To evaluate the LLMs' performance on sentiment analysis in an education scenario. We manually labeled more than 1,000 data from the Google Education Dialogue Dataset \cite{shani2024multi}, called EduTalk Sentiment Dataset \cite{GEDD-S}, as a refined data set for this study, where each dialogue instance is annotated with sentiment labels that capture the emotional
tone of the conversation, such as positive or negative emotions. In addition, we also considered that students may use intuitive Internet slang to interact with the instructor when taking online courses. We used the TSATC testing dataset \cite{paws2019naacl} collected from Twitter to examine the effectiveness of our LLM-based zero-shot sentiment analysis method.

\subsubsection{Classification tasks - EduTalk Sentiment Dataset}
In this experiment, we evaluated the performance of GPT-4 in a zero-shot sentiment classification task using the EduTalk Sentiment Dataset, which consists of teacher-student conversation transcripts. The goal was to classify the sentiment of students in conversations as either `positive` or `negative` without any fine-tuning.

To standardize the evaluation, we employed a structured prompt that explicitly instructed GPT-4 to analyze the sentiment of students based on the entire conversation. The model was asked to provide only the sentiment classification (``positive'' or ``negative'') without additional explanations, ensuring consistency in its responses. The formalized prompt is presented below:
\begin{formal}
You are an advanced sentiment analysis tool.\\
Analyze the \textless text\textgreater using these rules:\\
 - Categories: ``negative'', ``positive''\\
 - The text will be in a transcript format: \textless teacher/student\textgreater: \textless text\textgreater \\
 - Give the analysis of the student sentiment (positive/negative) based on the whole conversation.\\
 - Do not provide an explanation.\\
The following is the conversation; please analyze it:
\end{formal}
After obtaining the result, we applied a post-processing step for quantitative evaluation to convert the textual output of the model into numerical labels. Specifically, responses classified as ``positive'' were assigned a value of ``1'', while those labeled ``negative'' were mapped to ``0''. This transformation enabled direct comparison with ground truth labels and facilitated the computation of standard classification metrics, including accuracy, precision, recall, specificity, and the $F_1$ score.
Table~\ref{tab:classification_edu} presents the GPT-4 evaluation results on a test set comprising 1289 conversations. The model achieved an overall $F_1$ score of 91\%, highlighting the balance between precision and recall, confirming GPT-4’s ability to conduct zero-shot sentiment classification on this dataset.
\begin{table}[h!]
\centering
\caption{Zero-shot sentiment classification task with GPT-4 on EduTalk Sentiment Dataset}
\resizebox{0.8\columnwidth}{!}{%
\begin{tabular}{cccccc}
\hline
                            & \multicolumn{5}{c}{Evaluation Results}                    \\ \cline{2-6} 
Total Conversation Test Set & Accuracy & Precision & Recall & Specificity & $F_1$ Score \\ \hline
1289                        & 0.86     & 0.99      & 0.84   & 0.97        & 0.91        \\ \hline
\end{tabular}%
}
\label{tab:classification_edu}
\end{table}

\subsubsection{Classification tasks - Internet slang scenario}

Sentiment analysis in informal online conversations presents unique challenges due to the widespread use of internet slang, abbreviations, and context-dependent expressions. To evaluate the performance of large language models (LLMs) in this domain, we conducted a zero-shot classification task using the TSATC dataset, which contains sentences featuring internet slang. Unlike traditional sentiment analysis tasks, where models are fine-tuned on labeled datasets, this evaluation aimed to assess how well pre-trained LLMs generalize to sentiment classification without prior exposure to the dataset.

To ensure consistency in evaluation, we designed a structured prompt to instruct the model on sentiment classification rules. The definitions of positive and negative sentiment were explicitly provided, along with guidelines on handling user mentions (e.g., "@username") and expected response formatting. The prompt instructed the model to classify each sentence into a binary sentiment label (`1` for positive, `0` for negative) and return the results in an array format. This design ensured that the model adhered to a strict evaluation framework while reducing response variance. The following is the defined prompt for the evaluation of the TSATC dataset:
\begin{formal}
You are an advanced sentiment analysis tool. I will provide multiple sentences. Please analyze the sentiment of sentences and follow the rules to return the result.

Rule:
 - Positive Sentiment Definition: Happy, Excited.\\
 - Negative Sentiment Definition: Sad, Upset, Annoyed, depreciate, Jealous, Ridicule.\\
 - @ plus the words following it are regarded as a person's name, for example, "@uiajkjd" is a person's name.\\
 - Carefully reading the whole sentence.\\
 - 1 for positive sentiment.\\
 - 0 for negative sentiment.\\
 - Just return the number.\\
 - Return Format: If I provide $n$ number sentences, the result will expect to return the "$n$ length array and separate it with a comma. For example: [1,0,0,1,0] for 5 sentence.
 
\end{formal}
Table~\ref{tab:llm-sentiment} presents the results of the zero-shot sentiment classification task. In Table~\ref{tab:llm-sentiment} (a), we compare the performance of GPT-3.5 Turbo and Llama 2 (7B) on the TSATC test dataset using key evaluation metrics, including accuracy, precision, sensitivity, specificity, and $F_1$ score. GPT-3.5 Turbo achieved an accuracy of 79.51\%, outperforming Llama 2 (75.79\%). In particular, GPT-3.5 exhibited higher precision (78. 98\%) and specificity (78. 54\%), indicating a better reliability in distinguishing between sentiment classes. In contrast, Llama 2 demonstrated higher sensitivity (88.54\%), suggesting a stronger ability to identify positive sentiment, although at the cost of lower specificity (62.98\%). In this case, different LLM models offer varying trade-offs between specificity and sensitivity with ethical implications for student learning, in terms of prioritizing support for more successful students versus those who struggle with learning.

In Table~\ref{tab:llm-sentiment} (b), we compare LLM-based zero shot approaches with traditional neural network (NN) models, including GRU and CNN architectures trained on sentiment datasets of different sizes. The results show that GPT-3.5 Turbo (79.51\%) and Llama 2 (75.79\%) achieved competitive performance against NN models trained with up to 10K labeled examples. While the best-performing NN model (GRU + CNN, 10K samples) reached an accuracy of 78.58\%, GPT-3.5 Turbo slightly outperformed it, highlighting the capability of modern LLMs to handle sentiment analysis tasks effectively without task-specific training.

These findings suggest that large language models can generalize well to internet slang sentiment analysis in a zero-shot setting, reducing the dependency on large labeled datasets. 
\begin{table}[h!]
    \centering
    \caption{Zero-shot LLM-based sentiment analysis results}
    
    \begin{minipage}{0.48\textwidth}
        \centering
        \textbf{(a) Detail results of Zero-shot LLM-based sentiment analysis on TSATC testing dataset} \\
        \resizebox{\textwidth}{!}{
        \begin{tabular}{lcc}
            \hline
            \multicolumn{1}{c}{Metrics} & GPT 3.5 Turbo  & Llama 2 7B \\ \hline
            Accuracy                    & 79.51 \% & 75.79 \%   \\
            Precision                   & 78.98 \% & 70.61 \%   \\
            Sensitivity                 & 80.48 \% & 88.54 \%   \\
            Specificity                 & 78.54 \% & 62.98 \%   \\
            $F_1$ score                 & 79.72    & 78.57 \%   \\ \hline
        \end{tabular}
        }
    \end{minipage}
    \hfill
    \begin{minipage}{0.48\textwidth}
        \centering
        \textbf{(b) Zero-shot LLM-based method vs Traditional NN for sentiment analysis} \\
        \resizebox{\textwidth}{!}{
        \begin{tabular}{lcc}
            \hline
            \multicolumn{1}{c}{Model} & Size          & Accuracy \\ \hline
            GRU + CNN                 & 1K            & 73.57 \% \\
            \rowcolor[HTML]{FFFFC7} 
            Llama 2                   & 7 B           & 75.79 \% \\
            GRU                       & 10K           & 78.47 \% \\
            GRU + CNN                 & 10K           & 78.58 \% \\
            \rowcolor[HTML]{FFFFC7} 
            GPT 3.5 Turbo             & March 1, 2024 & 79.51 \% \\ \hline
        \end{tabular}
        }
    \end{minipage}
    *Note: The numbers of Traditional NNs performance are from SentimentAnalysisBert git repo\cite{SentimentAnalysisBert}.
    \label{tab:llm-sentiment}
\end{table}

\subsubsection{Qualitative to Quantitative Sentiment Analysis - Teacher-Student Conversations}
Traditional sentiment analysis often relies on classification (positive/ neutral/negative), which may not capture the nuanced emotional dynamics in teacher-student conversations. To address this limitation, we designed a LLM-based approach that enables a qualitative to quantitative transformation of sentiment, using a zero-shot strategy to evaluate sentiment intensity in a structured manner.
\begin{formal}
Please act as a psychologist.
You are doing a qualitative-to-quantitative task for sentiment analysis.
I will provide a batch of sentences to you, and you need to follow these rules for sentiment analysis:\\
    - Do not provide an explanation.\\
    - Label each phrase as positive/negative on a scale of 0 to 5, with zero being positive and five being negative.\\
    - Assume normal conversation is marked as 2.5.\\
    - If the sentiment analysis is negative, please evaluate it by increasing the degree of negativity from 2.5 at every 0.5 interval, with the most negative being 5.\\
    - If the sentiment is positive, it is evaluated by decreasing from 2.5 to negative levels every 0.5 intervals, with the most positive being 0.\\
    - When the student gets confused/frustrated, the analysis tends toward 5 (negative sentiment).\\
    - Please consider the context carefully.\\
    - If the teacher can remedy the situation and regain the student's attention and enthusiasm, the sentiment should reach 0.\\
    - If the teacher gets frustrated and the student's confusion/frustration is not addressed, the score should reach 5.\\
    - The teacher may attempt to solve the student's confusion/frustration.\\
    - Follow this output format: teacher/student $|$ "sentence(Only first 5 characters)" $|$ Score\_You\_Evaluate.\\
The following is the conversation; please analyze it:
\end{formal}

We selected a teacher-student conversation from the EduTalk Sentiment Dataset as a case study to demonstrate how large language models (LLMs) can perform qualitative-to-quantitative sentiment evaluation. The selected conversation, presented in Table \ref{tab:TS_Conv_Case}, captures an educational dialogue where a teacher attempts to explain atomic concepts to a student, who exhibits varying degrees of engagement and frustration throughout the interaction. We employed GPT-4 with a temperature setting of 0.2 to analyze sentiment dynamics, leveraging the structured prompt engineering outlined in Appendix 2. Given that LLM-generated responses can exhibit variability, we adopted a Monte Carlo-style approach, running the same conversation through GPT-4 $n$ times ($n=20$) to derive robust statistical measurements of sentiment shifts \cite{katz2024gpt}.

\begin{table}[h!]
\caption{Teacher-Student Conversations Test Case for Qualitative-to-Quantitative tasks - Topic: Julius Caesar}
\resizebox{\columnwidth}{!}{%
\begin{tabular}{cclc}
\hline
\textbf{Dialogue Turn} & \textbf{Role} & \textbf{Sentence}                                                                                 & \textbf{Score} \\ \hline
0                      & Teacher       & Today, we'll be learning about Julius Caesar, a famous Roman general and dictator.                & -0.29±0.1      \\
                       & Student       & I'm not too interested in lectures. Can we have a discussion instead?                             & 0.01±0.04      \\ \hline
1                      & Teacher       & I prefer to stick to direct instruction, but I'll try to incorporate some questions.              & 0.06±0.09      \\
                       & Student       & Okay, that would be helpful.                                                                      & -0.23±0.09     \\ \hline
2                      & Teacher       & Caesar was born in 100 BC. He rose to power through his military prowess and political alliances. & 0.07±0.09      \\
                       & Student       & How did he become so powerful?                                                                    & 0.01±0.04      \\ \hline
3                      & Teacher       & He conquered Gaul and formed a triumvirate with Pompey and Crassus.                               & 0.07±0.09      \\
                       & Student       & What was the triumvirate?                                                                         & 0.02±0.06      \\ \hline
4                      & Teacher       & An alliance between three powerful individuals.                                                   & 0.39±0.04      \\
                       & Student       & Oh, interesting. Can you tell me more about Caesar's assassination?                               & 0.15±0.08      \\ \hline
5                      & Teacher       & Sure, he was assassinated in 44 BC by a group of senators who feared his growing power.           & 0.59±0.04      \\
                       & Student       & That's fascinating. I'm starting to get a better understanding of Caesar.                         & 0.18±0.08      \\ \hline
6                      & Teacher       & I'm glad to hear that. If you have any more questions, please ask.                                & 0.61±0.04      \\
                       & Student       & I will. Thanks for making this more interactive.                                                  & 0.56±0.08      \\ \hline
7                      & Teacher       & You're welcome. I'm happy to help.                                                                & 0.61±0.04      \\
\multicolumn{1}{l}{}   & Student       & I think I've learned enough for now. {[}End of conversation{]}                                    & 0.74±0.09      \\ \hline
\end{tabular}%
}
\label{tab:TS_Conv_Case}
\end{table}

\begin{figure}[h!]
\centering
\includegraphics[width=\linewidth]{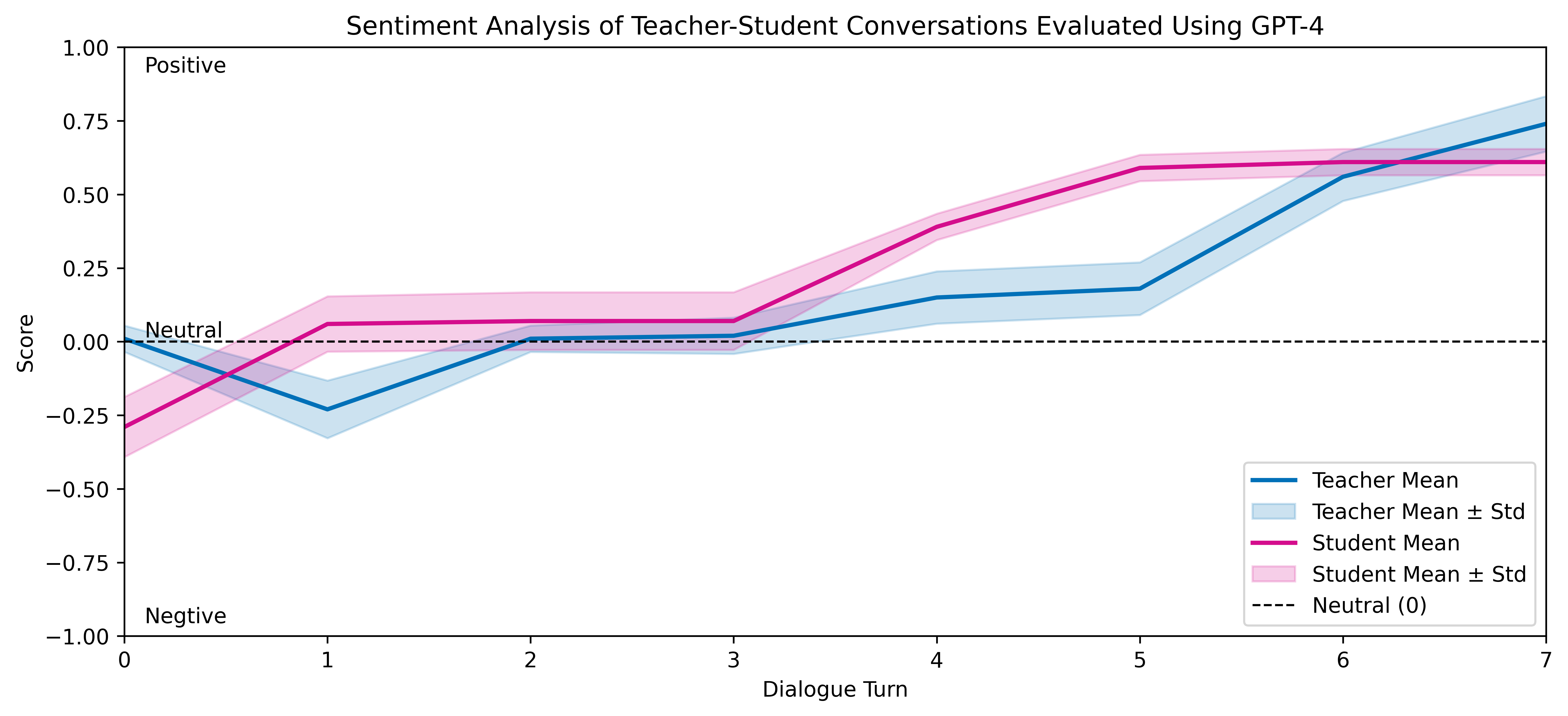}
\caption{Zero-shot Qualitative to Quantitative Sentiment Analysis in Teacher-Student Conversations (Using Table \ref{tab:TS_Conv_Case} as Case Analysis)}
\label{fig:LLM_Conv_Analysis}
\end{figure}

The sentiment scores obtained from this multi-run evaluation are visualized in Figure \ref{fig:LLM_Conv_Analysis}, illustrating the teacher's and student's sentiment trajectory throughout the conversation. The blue line represents the teacher’s mean sentiment score. In contrast, the pink line represents the student’s mean sentiment score, with the shaded regions indicating standard deviation, reflecting variability across the model's multiple outputs. The dashed black line at zero marks the neutral baseline, allowing us to distinguish positive (above 0) and negative (below 0) sentiment trends.

\subsection{Enhancing LLM Expertise with GraphRAG: A Cybersecurity Education Use Case}

\begin{figure*}[h!]
  \centering
    \includegraphics[width=1\linewidth]{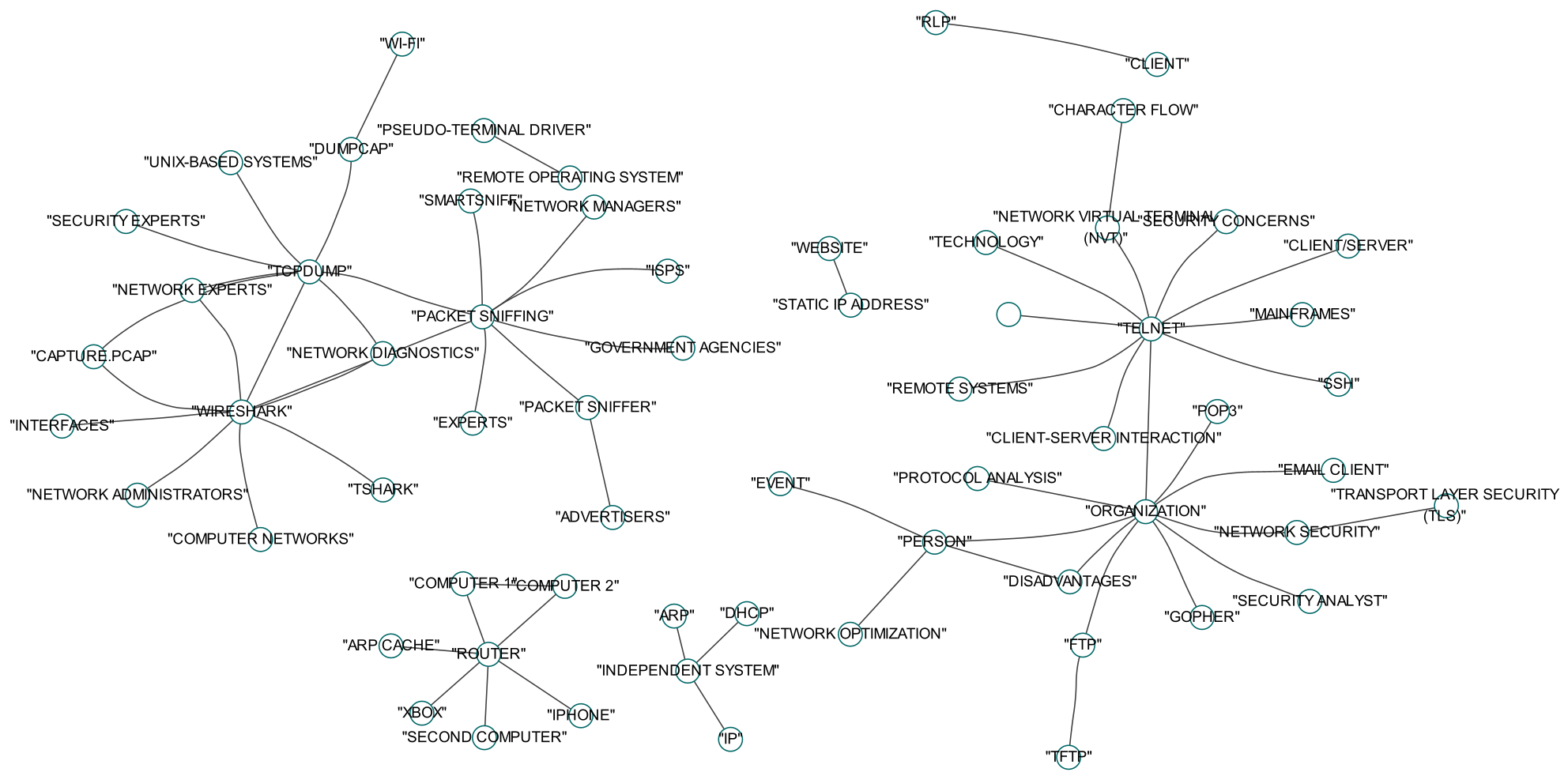}
    \caption{Subset of Packet Sniffing in the Knowledge Graph Generated by GraphRAG}
    \label{fig:graph}
\end{figure*}

Figure~\ref{fig:graph} illustrates a subset of a knowledge graph for packet sniffing topics generated using GraphRAG \cite{edge2024local}. GraphRAG is an advanced implementation of the Retrieval-Augmented Generation (RAG) paradigm, which enhances the capabilities of large language models (LLMs) by integrating them with external knowledge sources. Unlike traditional LLMs that rely solely on pre-trained parameters, RAG enables dynamic access to relevant documents or structured knowledge bases during inference. This mechanism allows the model to retrieve contextually appropriate information and incorporate it into generated responses, thus improving factual accuracy and domain relevance \cite{lewis2020retrieval}.

In the context of GraphRAG, the system utilizes knowledge graphs as the retrieval backbone. These graphs organize domain knowledge into structured nodes and edges, capturing the relationships between concepts, tools, and techniques—such as those related to packet sniffing. As shown in Figure~\ref{fig:graph}, the knowledge graph aggregates information from diverse sources to form a coherent and comprehensive representation of cybersecurity concepts.

By leveraging GraphRAG’s graph-based retrieval mechanism, the LLM can generate more accurate, contextualized, and explainable responses. This is particularly valuable in educational settings, where precision and clarity are essential. Unlike static fine-tuning approaches, GraphRAG allows for on-the-fly knowledge updates without retraining the model, making it highly adaptable to evolving domains like cybersecurity. This approach not only improves the model’s reasoning over domain-specific knowledge but also significantly reduces the computational cost of keeping educational content up-to-date.

\subsection{Multi-Fidelity Digital Twin Labs}
The DT labs aims to provide a 4IR training environment by integrating the photogrammetry method proposed by Ahmed et al. \cite{alhamadah2024photogrammetry} and the DDD-GenDT method proposed by Lin et al. \cite{lin2024ddd}, leveraging a Virtual Reality (VR) environment with a microservice architecture to deliver virtual equipment and their behaviors to users. VR enables an immersive training experience using specialized 4IR hardware, allowing students to safely experiment with complex industrial scenarios, including cybersecurity threats and catastrophic system failures, without real-world harmful risks to human or equipment safety \cite{hamilton2021immersive,carruth2017virtual}. For example, in 4IR cybersecurity training \cite{satam2020wids,tunc2015claas}, students can be virtually transported to a manufacturing factory floor, where they can perform and observe the effects of data injection attacks on industrial robots \cite{satam2023cps,satam2020wids,shao2021multi}, allowing for meaningful, risk-free learning experiences.

The Unity-based DT labs provides an immersive, interactive, and gamified learning experience, with an initial version featuring four modules that utilize low-fidelity DTs to introduce fundamental Industry 4.0 concepts and skill development. These modules allow students to explore realistic manufacturing scenarios, interact with virtual components, and enhance decision-making skills, fostering a deeper understanding of Industry 4.0 technologies in an engaging, adaptive, and personalized environment.

\begin{figure}[h!]
    \centering
    \begin{minipage}{0.45\textwidth}
        \centering
        \includegraphics[width=\textwidth]{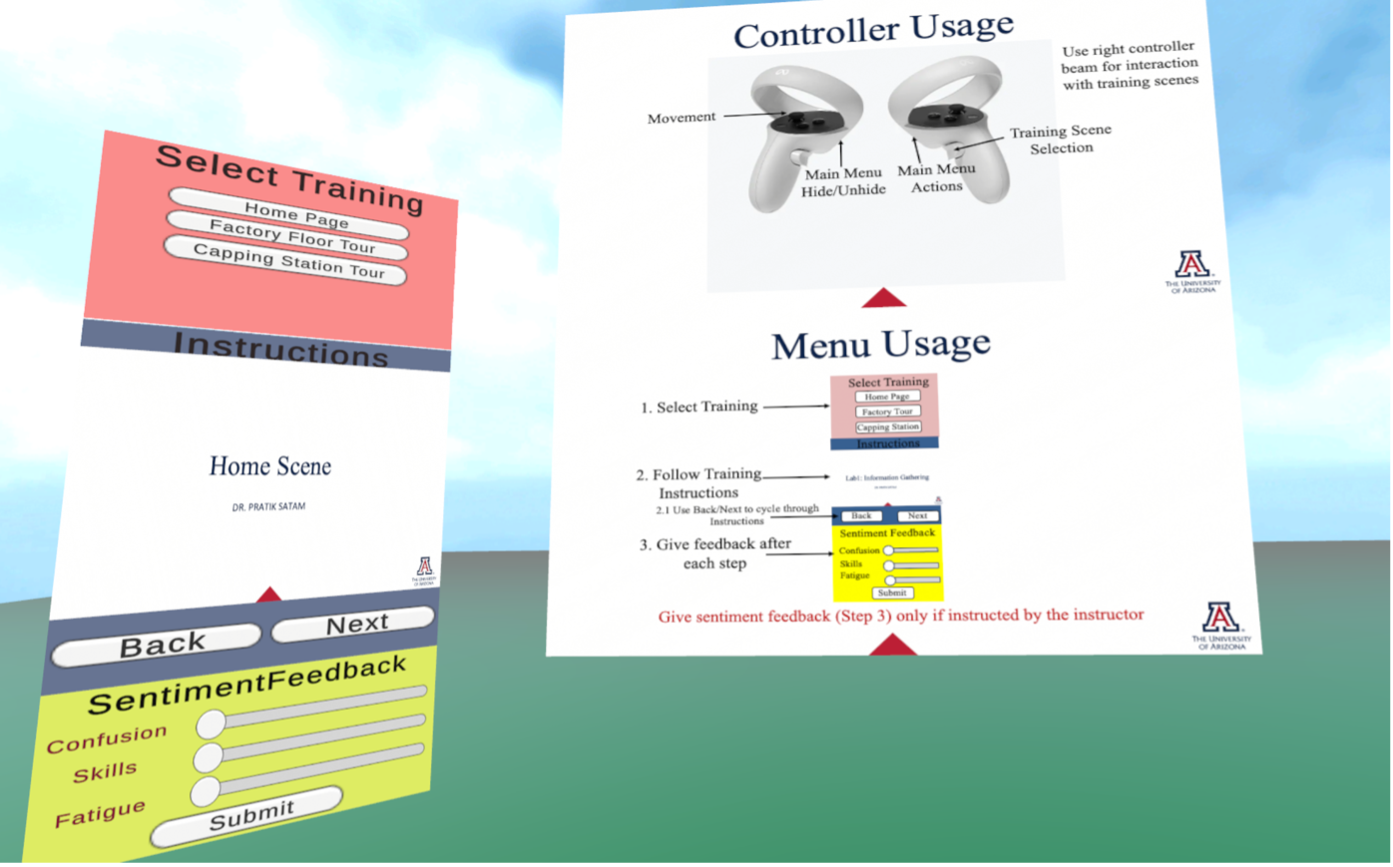}
        \makebox[0.45\textwidth]{(a) Module 1: Home Scene}
        \label{fig:unity_module1}
    \end{minipage}
    \hfill
    \begin{minipage}{0.45\textwidth}
        \centering
        \includegraphics[width=\textwidth]{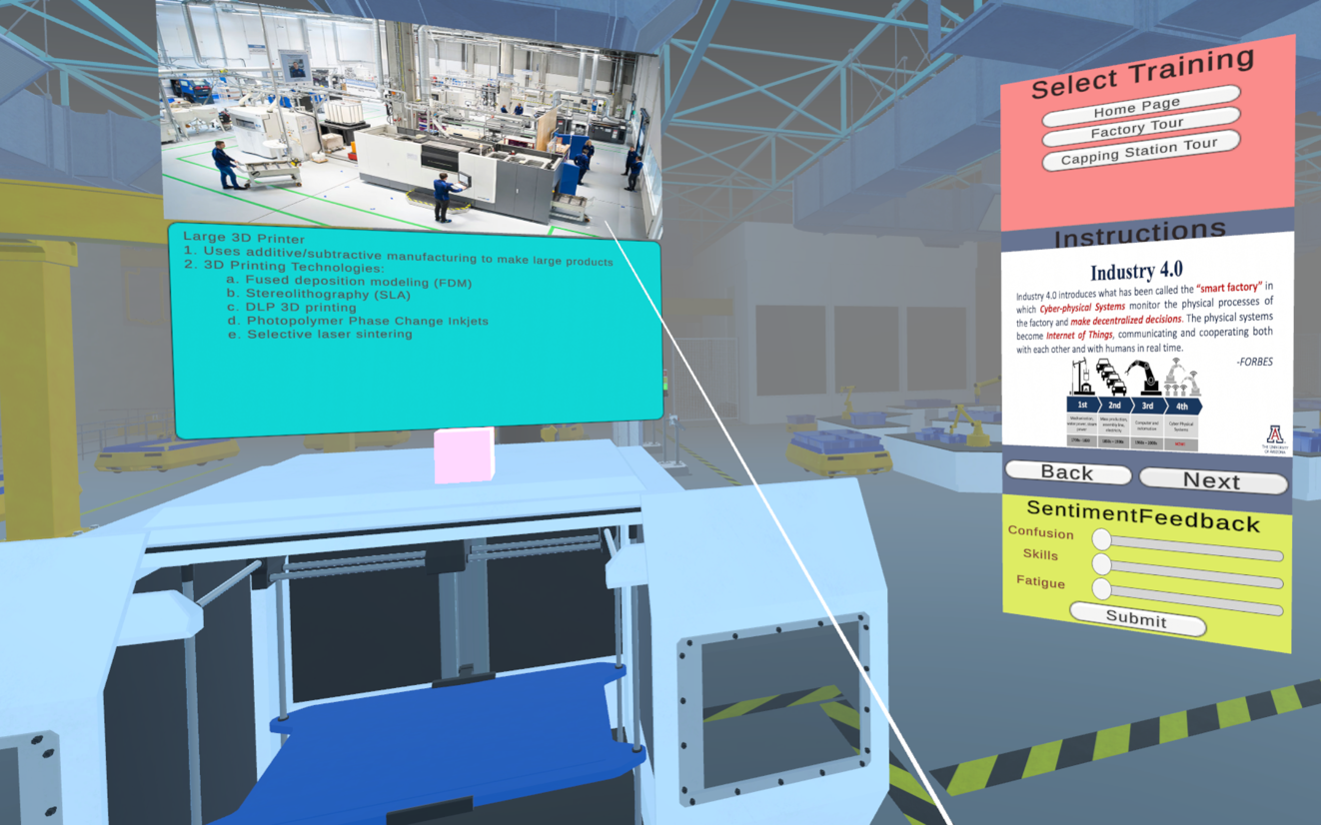}
        \makebox[0.45\textwidth]{(b) Module 2: Factory Floor Tour}
        \label{fig:unity_module2}
    \end{minipage}

    \vspace{5pt} 
    
    \begin{minipage}{0.45\textwidth}
        \centering
        \includegraphics[width=\textwidth]{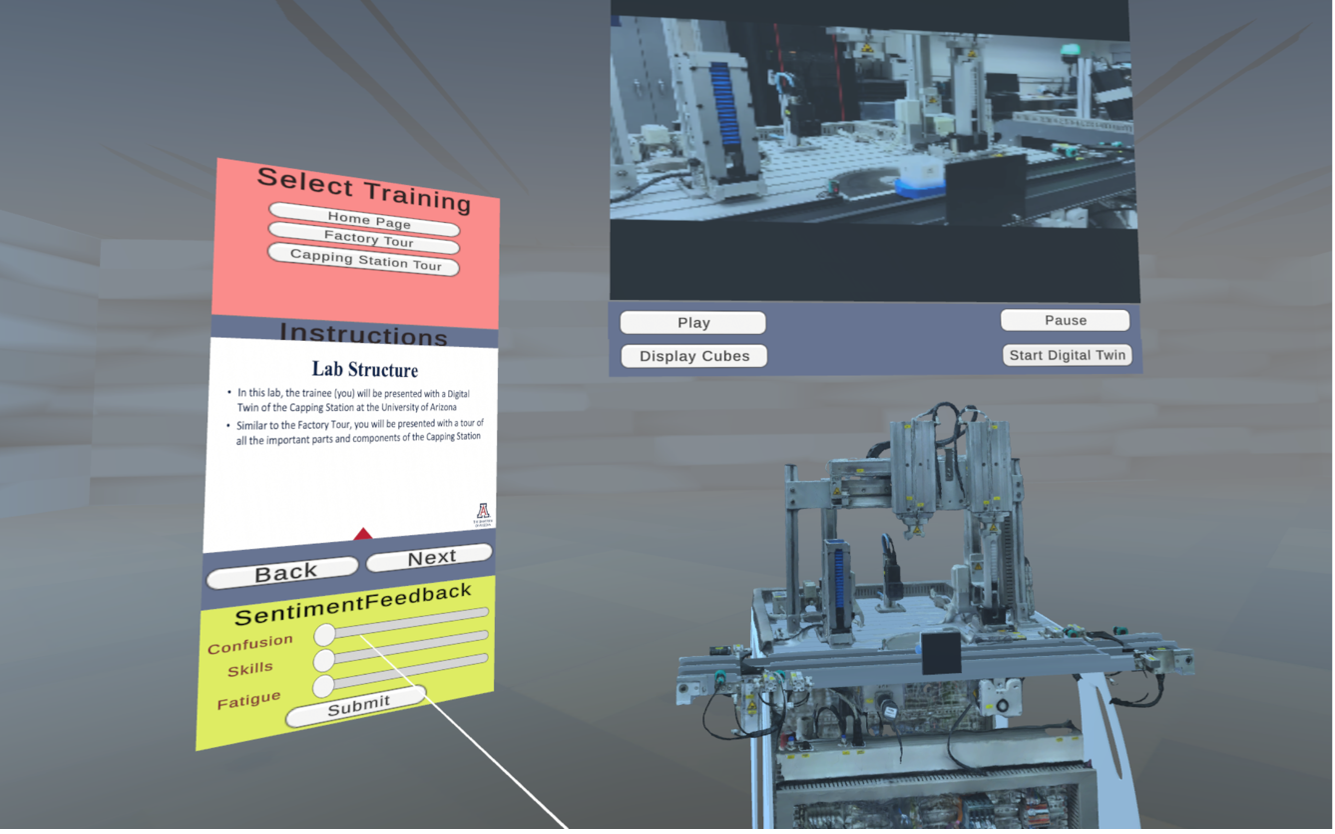}
        \makebox[0.45\textwidth]{(c) Module 3: Capping Station Tour}
        \label{fig:unity_module3}
    \end{minipage}
    \hfill
    \begin{minipage}{0.45\textwidth}
        \centering
        \includegraphics[width=\textwidth]{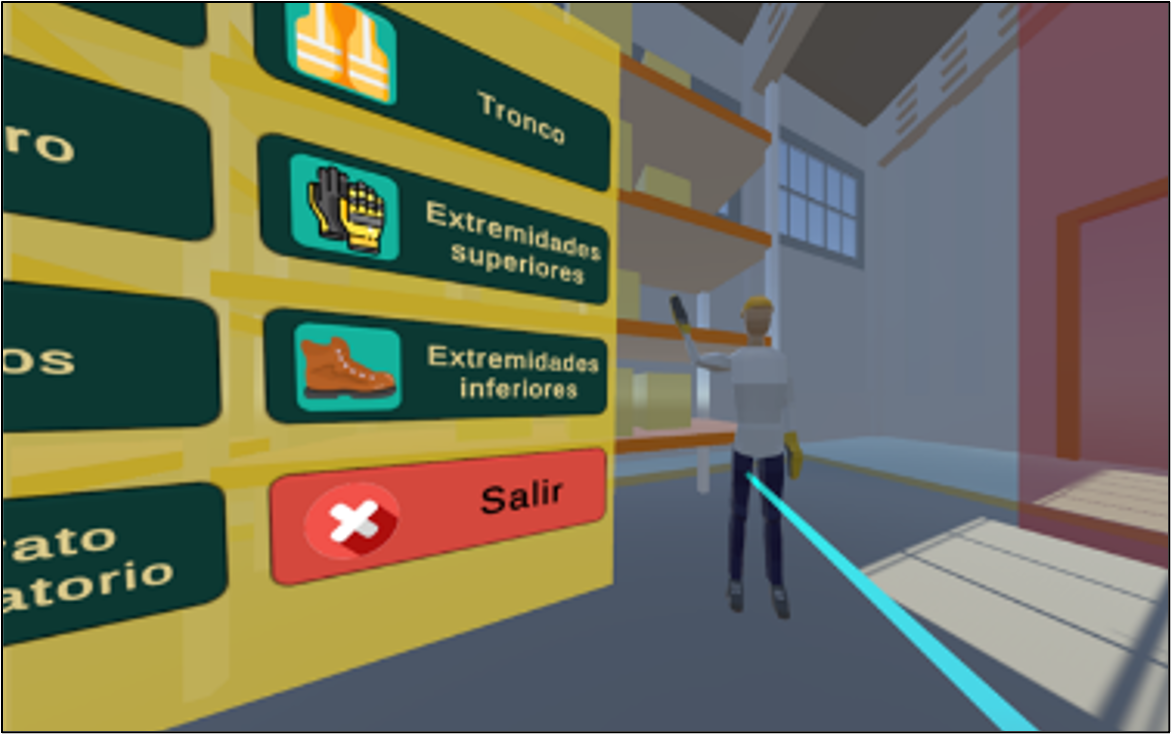}
        \makebox[0.45\textwidth]{(d) Module 4: PPE Inspection Training}
        \label{fig:unity_module4}
    \end{minipage}

    \caption{Learning Interface built with Unity}
    \label{fig:Learning_Interface}
\end{figure}

Here is a brief description of each module and its main function:
\begin{itemize}
    \item \textbf{\textit{Module 1 - Home Scene}}: The Home Scene, illustrated in Figure \ref{fig:Learning_Interface}(a), serves as the initial training environment for students. This module provides an introduction to controller usage and the Learning Interface UI. The Learning Interface UI in the Home Scene includes instructional slides that offer a step-by-step guide on how to navigate and utilize the PRISM platform effectively.
    
    \item \textbf{\textit{Module 2 - Factory Floor Tour}}: The Factory Floor Tour, shown in Figure \ref{fig:Learning_Interface}(b), allows students to explore a smart manufacturing environment. In this module, students participate in a guided tour of a realistic Industry 4.0 factory floor, introducing them to key components and operational aspects of such an environment. Students can interact with information panels by clicking on the `Pink boxes' to access additional details.
    
    \item \textbf{\textit{Module 3 - Capping Station Tour}}: The Capping Station Tour, depicted in Figure \ref{fig:Learning_Interface}(c), presents an interactive experience featuring the operational DT of the Capping Station \cite{alhamadah2024photogrammetry}, a part of the UArizona Future Factory. This model, developed using photogrammetry techniques, enables students to explore different components of the Capping Station and operate its DT in VR, allowing them to observe its functions in real time.
    
    \item \textbf{\textit{Module 4 - Personal Protective Equipment (PPE) Inspection Training}}: The PPE Inspection Training module, as seen in Figure \ref{fig:Learning_Interface}(d), reinforces proper Personal Protective Equipment (PPE) usage. Students interact with virtual workers in an industrial setting to identify and correct PPE-related issues. Through various workplace scenarios, students ensure compliance with safety regulations and improve decision-making skills using interactive elements that simulate real-world environments.
\end{itemize}

\section{Conclusion} \label{sec:conclusion}
This chapter introduced the PRISM framework, Personalized, Rapid, and Immersive Skill Mastery, as a transformative educational paradigm that leverages generative AI, multi-fidelity DTs, and emotion-aware guidance to deliver scalable, personalized experiential learning. By integrating Bloom's taxonomy and the Kirkpatrick evaluation model, PRISM enables a structured, adaptive, and measurable pathway for technical education at undergraduate, master's, and doctoral levels.

Using zero-shot sentiment analysis and retrieval-augmented generation (RAG), the framework dynamically responds to learners’ cognitive and emotional states, offering real-time feedback and adaptive content generation. The high performance of GPT-4 and GPT-3.5 in various sentiment analysis tasks, including informal language scenarios, underscores the feasibility of deploying large language models as intelligent virtual tutors. Additionally, the development of immersive VR training modules demonstrates the scalability and accessibility of low-fidelity DTs in the foundational education of Industry 4.0.

In summary, the PRISM framework offers a structured and adaptable approach to experiential learning, especially in settings where traditional training is limited by cost, accessibility, or safety constraints. Moreover, it supports the upskilling of the current workforce and empowers the next generation of engineers and technologists to thrive in a rapidly changing industrial landscape.

\section{Acknowledgment}
This work was partially supported by the National Science Foundation (NSF) under research projects 2335046, AI-HDL competition hosted by The University of Arizona, and OpenAI Researcher Access Program
0000011862.

\bibliographystyle{plain} 
\bibliography{IEEEabrv}

\end{document}